\documentclass[aps,prl,twocolumn,superscriptaddress]{revtex4}
\usepackage{epsf,graphicx}
\usepackage{amssymb}
\usepackage{amsmath}
\usepackage{latexsym,bm,array,amsfonts,multirow}
\usepackage{color}
\usepackage{ulem}
\begin{document}
	
\title{Chiral finite-momentum superconductivity in the tetralayer graphene}
	\author{Qiong Qin}
	\affiliation{New Cornerstone Science Laboratory, Department of Physics, School of Science, Westlake University, Hangzhou 310024, Zhejiang, China}
	\author{Congjun Wu}
	\email[]{wucongjun@westlake.edu.cn}
	\affiliation{New Cornerstone Science Laboratory, Department of Physics, School of Science, Westlake University, Hangzhou 310024, Zhejiang, China}
	\affiliation{Institute for Theoretical Sciences, Westlake University, Hangzhou 310024, Zhejiang, China}
	\affiliation{Key Laboratory for Quantum Materials of Zhejiang Province, School of Science, Westlake University, Hangzhou 310024, Zhejiang, China}
	\affiliation{Institute of Natural Sciences, Westlake Institute for Advanced Study, Hangzhou 310024, Zhejiang, China}
	\begin{abstract}
	Motivated by the recent experimental discovery of superconductivity in rhombohedral tetralayer graphene, we investigate the pairing mechanism arising from the density-density interactions within the random-phase approximation. 
    This approach successfully highlights the dominance of the chiral $p$-wave pairing between electrons with the same spin and valley index at low densities, while also predicting the superconducting range in agreement with experimental findings. 
    Furthermore, we examine the characteristics of distinct superconducting regions: SC1 and SC2 exhibit chiral finite-momentum superconductivity with pronounced phase fluctuations, whereas SC4 displays zero-momentum spin-singlet superconductivity. 
	\end{abstract}
	\maketitle
	
   \textit{Introduction.--} 
   Although superconductivity has been found in a variety of layered material systems \cite{Zhou2021c,Cao2018,Chen2019a,Hao2021,Park2021a,Cao2021a,Li2024,Xia2024,Guo2024,Choi2024}, the recent discovery of superconductivity in rhombohedral tetralayer graphene \cite{Han2024a} still generates renewed and significant interest \cite{Chou2024,Geier2024,Kim2024,Wang2024a}. 
   Such a system exhibits numerous exotic properties, including an anomalous Hall effect in the spin-valley-polarized metallic phase, as well as time-dependent resistivity fluctuations and magnetic hysteresis in the superconducting state \cite{Han2024a}. Additionally, the high upper critical field ($H_{c2}$) and the coherence length comparable to the inter-particle distance further underscore the unconventional nature of pairing.
   Nevertheless, a comprehensive understanding of these unusual properties, along with the underlying pairing mechanism, remains limited and warrants thorough investigation and detailed analysis.

    Meanwhile, the effective electron density in this system is notably low, approximately $0.5 \times 10^{12}$cm$^{-2}$, 
    suggesting the role of strong interaction in the pairing mechanism. 
    Moreover, the presence of an extremely flat band further inhibits the motion of Cooper pairs.
    Both aspects would lead to strong superconducting phase fluctuations, and consequentially small superfluid density.  
    An important question naturally arise: Do phase fluctuations impose a fundamental constraint on superconductivity across all superconducting regions in the tetralayer graphene? 
    Additionally, 
    although the nature of spin singlet and triplet pairings can be inferred by comparing 
    $H_{c2}$ and the Pauli limit, whether pairing occurs within the same valley remains to be clarified.
    
\begin{figure}[t]
\begin{center}
\includegraphics[width=8cm]{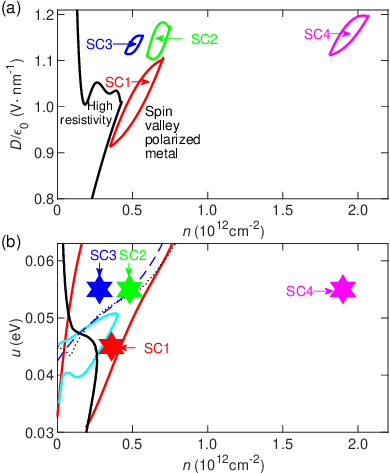}
\end{center}
\caption{Schematic (a) experimental \cite{Han2024a} and (b) theoretical phase diagram showing the electronic density $n$ ($10^{12}$cm$^{-2}$)  and the electric displacement field $D/\epsilon_0$ (V nm$^{-1}$) or $u$ (eV), where the shape of the low-$T_c$ constraint area in (b) is similar to that of the high-resistivity region in (a) and SC1–SC4 denote the different superconducting regions. We also note that the cyan lines denote the boundaries of  $r_s > 40$ region, where $r_s$ is the ratio of the interaction energy to the kinetic energy. The maximum of pairing gap (the dashed blue line) corresponds to peaks of density of state $N(E_F)$ ( the dotted black line) at Fermi energy.
}
\label{fig1}
\end{figure}

    Motivated by these observations and questions, we conduct a self-consistent mean-field calculation with the effective Coulomb interaction based on the random-phase approximation (RPA). 
    Upon identifying the dominance of the $p$-wave pairing, we examine the dependence of superconducting transition temperature ($T_c$) and the gap function on the electron density $n$ ($10^{12}$cm$^{-2}$) and the electric displacement field $u$ (eV).
    The theoretical analysis yields the regions of superconductivity aligning well with experimental observations. 
    When combined with superfluid density calculations, it highlights the impact of phase fluctuations across different regions. 
    While superconductivity at low electron densities appears constrained by phase fluctuations or limited electron mobility, $T_c$'s at higher densities seem predominantly governed by the pairing strength. 
    Additionally, we characterize the nature of superconducting phases SC1–4 in Fig.~\ref{fig1}, providing a basic understanding of the experimental phase diagram. 
    SC1 and SC2 correspond to a chiral finite-momentum intra-valley pairing with parallel spins, while SC4 exhibits the spin singlet pairing between opposite valleys carring zero momentum. 
    The pairing in SC3 is chiral, zero-momentum, and occurring between opposite valleys.

\textit{Method and Model.--} 
To describe the band structure of the rhombohedral tetralayer graphene, we adopt the following tight-binding model \cite{Zhou2021c,Chou2021a,Ghazaryan2021b,Qin2023c,Han2024a,Murshed2025} for the non-interacting part:
\begin{eqnarray}
	H_0=\sum_{\bm{k}s\tau}\Psi_{\bm{k}\tau s}^{\dagger}h_{\bm{k}\tau}\Psi_{\bm{k}\tau s}.
    \label{eq:H0}
\end{eqnarray}
$\tau$ and $s$ denote the valley and spin 
indices, respectively;
$\Psi_{\bm{k}\tau s}= (\psi_{\bm{k}\tau s,1A}, \cdots ,
\psi_{\bm{k}\tau s,4B})^T$, where $\psi_{\bm{k}\tau s,l\sigma}$ is the annihilation operator with the layer index $l=1\sim 4$  and the sublattice index $\sigma=A, B$. 
Without doping, four bands below zero energy are fully occupied, and the other four above zero are empty. 
See Supplemental Material (S. M.) I for details.

The density-density interaction is adopted, which are expressed in momentum space as:
\begin{eqnarray}
	H_I=\frac{1}{2A}\sum_{\bm{q}}V_{0,\bm{q}}\rho_{\bm{q}}\rho_{-\bm{q}},
    \label{eq:Hint0}
\end{eqnarray}
where the density operator is defined as $\rho_{\bm{q}}=\sum_{\bm{k}\tau s}\Psi_{\bm{k}\tau s}^{\dagger}\Psi_{\bm{k}+\bm{q}\tau s}$ and $V_{0,\bm{q}}=\frac{e^2}{2\epsilon q}\tanh(qd)$ represents the screened Coulomb interaction \cite{Ghazaryan2021b,Guerci2024,Kohn1965,Jimeno-pozo2023,Long2024}. 
In this context, $A$ is the 2D area of the system, $d=30$ (nm) is the distance between two gates and $\epsilon=\epsilon_0\epsilon_r$ is the dielectric constant with $\epsilon_r=5$.

Define $c_{\bm{k}\tau s,\mu}$  as the annihilation operator for each band $\mu=1\sim 8$ with the relation
$\psi_{\bm{k}\tau s,i}=\sum_{\mu}u_{\tau,i\mu}(\bm{k})c_{\bm{k}\tau s,\mu}$,
where $i=1\sim 8$ is the abbreviation of the layer and sublattice indices, 
and $u_{\tau,i\mu}(\bm{k})$ represents the matrix elements of diagonalizing $h_{\mathbf{k}\tau}$
in the band Hamiltonian Eq.~(\ref{eq:H0}).
Given that superconductivity occurs at very low doping, the Fermi energy lies in the first conduction band,
i.e., the band index $\mu_0=5$.
Therefore, the electron interaction Eq.~(\ref{eq:Hint0}) should be projected onto this band:
\begin{eqnarray}
H_I=\frac{1}{2A}\sum_{\bm{q}}V_{0,\bm{q}}\tilde{\rho}_{\bm{q}}\tilde{\rho}_{-\bm{q}},
\end{eqnarray}
in which $\tilde{\rho}_{\bm{q}}$ is the projected density operator defined as:
\begin{eqnarray}
	\tilde{\rho}_{\bm{q}}=\sum_{\bm{k}\tau s}\sum_i\left(u_{\tau,i\mu_0}^*(\bm{k})u_{\tau,i\mu_0}(\bm{k}+\bm{q})\right)c_{\bm{k}\tau s,\mu_0}^{\dagger}c_{\bm{k}+\bm{q}\tau s,\mu_0}.
    ~
\label{eq:prj_den}
\end{eqnarray}

Within the RPA approach, the renormalized interaction takes the form:
\begin{eqnarray}
	V_{\bm{q}}=\frac{V_{0,\bm{q}}}{1+\chi_{0,\bm{q}}V_{0,\bm{q}}},
\end{eqnarray}
where  $\chi_{0,\bm{q}}$ is the static susceptibility of the
projected density operator $\tilde \rho_{\mathbf{q}}$.
The detailed expression of $\chi_{0,\bm{q}}$ in the form of Lindhard response is given in S. M. II. 
The renormalized interaction can be represented as:
$H_I=\frac{1}{2A}\sum_{\bm{q}}V_{\bm{q}}\tilde{\rho}_{\bm{q}}\tilde{\rho}_{-\bm{q}}$.
Then the pairing interaction between electrons with the same spin and the same valley takes the following form: $H_I=\frac{1}{2A}\sum_{\bm{k}'\bm{k}\bm{Q}}V_{P,\bm{k}'-\bm{k}}c_{\bm{Q}/2+\bm{k}}^{\dagger}c_{\bm{Q}/2-\bm{k}}^{\dagger}c_{\bm{Q}/2-\bm{k}'}c_{\bm{Q}/2+\bm{k}'}$,
where
\begin{eqnarray}
    V_{P, \bm{k}'-\bm{k}}&=&V_{\bm{k}'-\bm{k}}\left[\sum_j\left(u_{j\mu_0}^*(\bm{Q}/2-\bm{k})u_{j\mu_0}(\bm{Q}/2-\bm{k}')\right)\right]
    \nonumber\\ 
   &\times& \left[\sum_{i}u_{i\mu_0}^*(\bm{Q}/2+\bm{k})u_{i\mu_0}(\bm{Q}/2+\bm{k}')\right].
\label{eq:Vkk}
\end{eqnarray}
We note that $V_{\bm{k}'-\bm{k}}$ is real, while the remaining part of $V_{P, \bm{k}'-\bm{k}}$ corresponds to the complex factor.

In calculations presented below, we first consider the intra-valley pairing by setting $\bm{Q}=2\bm{K}$ with $\bm{K}$ as the momentum of the Dirac point of the graphene Brillouin zone.
In the mean-field approximation, the gap equation reads:
\begin{eqnarray}
\Delta_{\bm{k},\bm{Q}} &=&-\frac{1}{A}
\sum_{\bm{k}'}
\frac{V_{P,\bm{k}'-\bm{k}}\Delta_{\bm{k}',\bm{Q}}}{4E_{\bm{k}',\bm{Q}}}\\
&\times&\left(\tanh \frac{\beta }{2} E_{\bm{k'},\bm{Q}}^{+}+
\tanh \frac{\beta }{2} E_{\bm{k'},\bm{Q}}^{-}\right), \nonumber
\end{eqnarray}
where $E_{\bm{k},\bm{Q}}
^\pm=\pm\frac{1}{2}\left(\xi_{\bm{Q}/2+\bm{k}}-\xi_{\bm{Q}/2-\bm{k}}\right)+E_{\bm{k},\bm{Q}}$,   
$E_{\bm{k},\bm{Q}}=\sqrt{\xi_{\bm{k},\bm{Q}}^2+
|\Delta_{\bm{k},\bm{Q}}|^2}$ and $\xi_{\bm{k},\bm{Q}}=\frac{1}{2}\left(\xi_{\bm{Q}/2+\bm{k}}+\xi_{\bm{Q}/2-\bm{k}}\right)$. Here, $\xi_{\bm{Q}\pm\bm{k}}$ is the conduction-band dispersion, and $\beta=1/T$ is the inverse temperature.


\begin{figure}[t]
\begin{center}
\includegraphics[width=8cm]{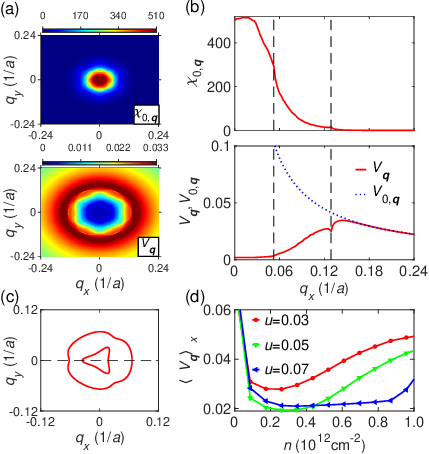}
\end{center}
\caption{(a) Momentum dependence of the static charge susceptibility $\chi_{0,\bm{q}}$ and interaction $V_{\bm{q}}$ at a displacement field of $u=0.055$ eV and an electron density of $n=0.5\times 10^{12}$ cm$^{-2}$.  
(b) Momentum dependence of $\chi_{0,\bm{q}}$, $V_{\bm{q}}$ and $V_{0,\bm{q}}$ along the $q_x$ direction.  
(c) The corresponding Fermi surface.  
(d) Distribution of the average interaction strength $\langle V_{\bm{q}}\rangle_x$ along the $q_x$ direction across different displacement fields $u$ and electron densities $n$.  
}
\label{fig2}
\end{figure}
 
\textit{Results.--}  
For simplicity, we initially ignore the complex factor in the projected pairing interaction $V_{P, \bm{k}'-\bm{k}}$, reducing its momentum dependence to $V_{\bm{k}'-\bm{k}}$, with $\bm{q}=\bm{k}'-\bm{k}$. 
Figure~\ref{fig2}(a) presents the momentum dependence of both the interaction $V_{\bm{q}}$ and the static charge susceptibility $\chi_{0,\bm{q}}$. 
As the momentum magnitude $q$ increases, 
$V_{\bm{q}}$ first rises and then drops, while $\chi_{0,\bm{q}}$
exhibits a monotonic decline as the bare interaction $V_{0,\bm{q}}$ does. 
To capture their detailed relations, 
we analyze their momentum dependence along the $q_x$ direction in Fig.~\ref{fig2} (b). 
Notably, the singularity in $V_{\bm{q}}$ arises from the scattering between two Fermi wavevectors along the direction of $q_x$, as illustrated in Fig.~\ref{fig2}(c), which shows the corresponding Fermi surfaces. 

The system is controlled by electron density $n$ ($10^{12} 
 ~$cm$^{-2}$) and 
electric displacement fields $u$ (eV).
Although $V_{\bm{q}}$ exhibits the similar momentum dependence pattern 
over different values of $n$ and $u$,
the distributions of its maximum and average values vary.
The average value of $V_{\bm{q}}$ along the $q_x$ direction 
is denoted as $\langle V_{\bm{q}}\rangle_x$, whose dependence
on $u$ and $n$ is depicted in Fig.~\ref{fig2}(d).
It shows that  $\langle V_{\bm{q}}\rangle_x$ decreases with increasing $u$, and as increasing $n$, $\langle V_{\bm{q}}\rangle_x$ initially declines before rising again. 
As shown in the schematic experimental phase diagram \cite{Han2024a} in Fig. ~\ref{fig1}(a), the electron density range for SC3 and SC2 are similar to that of SC1, which are around $n=0.2\sim0.8$ ($10^{12}$ cm$^{-2}$), but their corresponding displacement fields are significantly higher than those of SC1. 
On the other hand, the electron density in SC4 is much higher, approximately  $n=1.9$ ($10^{12}$ cm$^{-2}$). 
These observations suggest that SC4 and SC1 may have substantially stronger interaction strengths than those in SC2 and SC3. However, a crucial question remains: Do these enhanced interactions necessarily
results in a stronger pairing interaction and an increased pairing strength?  

\begin{figure}[t]
	\begin{center}
	\includegraphics[width=8cm]{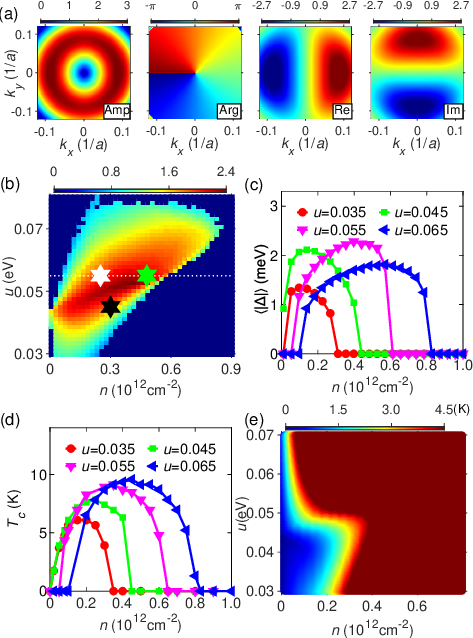}
	\end{center}
	\caption{(a) Momentum dependence of the pairing gap $\Delta_{\bm{k},\bm{Q}}$ at the electron density $n=0.5
    ~(10^{-12}$ cm$^{-2}$)
    with the displacement field $u=0.055$ (eV) for its amplitude (Amp), angle (Arg), real (Re) and imaginary (Im) component. 
    (b) Distribution of the average absolute value of gap  
    $\langle|\Delta|\rangle$ over different values of $n$ and $u$. The black, green, and white hexagrams respectively indicate the possible SC1–SC3 regions. The dotted white line marks the trajectory plotted in Fig.~\ref{fig4} and Fig.~\ref{fig5}.
    (c) $\langle|\Delta|\rangle$ as a function of $n$ at different values of $u$. 
    (d) Variation of $T_c$ with $n$ at different values of $u$. 
    (e) The distribution of the constraint of $T_c$ (K) over different values of $u$ and $n$. 
    We note that $\langle|\Delta|\rangle$ is evaluated at the low temperature of $T=0.001$ (K).
}
	\label{fig3}
\end{figure}

Given that SC1 and SC2 are situated near the spin–valley–polarized metallic phase, we first consider the intra-valley pairing with a center-of-mass momentum of $\bm{Q}=2\bm{K}$. The self-consistent mean-field calculations to the gap function are performed.
The solution converges to a state whose gap function $\Delta_{\bm{k},\bm{Q}=2\bm{K}}$ exhibits $p_{x}+ip_{y}$-wave symmetry.
In Fig.~\ref{fig3}(a), the amplitude (Amp), phase (Arg),
real (Re) and imaginary (Im) components of the gap function are displayed with the parameter values of $u=0.055$ (eV) and $n=0.5$ ($10^{12}$ cm$^{-2}$).
We define the average value of the gap function 
as $\langle|\Delta| \rangle = N_s^{-1}\sum_{\bm{k}} |\Delta_{\bm{k}, 2\bm{K}}|$, which is plotted in Fig.~\ref{fig3} (b) as a function of $n$ and $u$ at low temperature $T=0.001$ (K). Here, the cut off in momentum space is set inside the range $k_x, k_y \in [-0.12,0.12]$, and $N_s=84\times 84$ is the number of $\bm{k}$ points in the region we calculated.
Superconductivity is absent in most regions except in the vicinity of the maximum value of density of states (DOS) at Fermi energy $N(E_F)$ (see S. M. I), underscoring its sensitivity to DOS.
The superconducting region appears at intermediate values of both $u$ and $n$, whose overall shape resembles the experimental observations, although experimentally the superconducting regions SC1, SC2, and SC3 are disconnected. 
Figures~\ref{fig3}(c) and \ref{fig3}(d) illustrate the dependence of $\langle|\Delta|\rangle $ and 
$T_c$ on $n$ at different values of $u$. 
As $u$ increases, the range of the superconducting dome in terms of $n$ first expands and then shrinks. 
Moreover, both the starting and ending points of the superconducting dome are shifted rightward, i.e.,  to higher values. 
Notably, the maxima of $T_c$ or $\langle |\Delta|\rangle$ coincide with the peaks of the DOS $N(E_F)$ in the normal state, as shown in Fig.~\ref{fig1}(b).
 

Our calculation shows that superconductivity vanishes at $n > 0.8 ~(10^{-12}$ cm$^{-2})$, which is consistent with experiments.
Nevertheless, the calculated maximum of $T_c$ is much higher, reaching about $10$~K, compared to the experimental value of $0.3$~K~\cite{Han2024a}. 
Furthermore, experimentally superconductivity disappears at $n<0.2~(10^{-12}$ cm$^{-2})$, while our calculation show relatively high values of $T_c$.
This is due to the drawback of mean-field theory, which only captures the pairing strength but overlooks phase fluctuations. 
The actual $T_c$ is determined by the phase-coherence temperature, which is considerably lower.
Strong phase fluctuations at low electron density can even completely suppress superconductivity. 

Below we qualitatively estimate the suppression of $T_c$ due to phase fluctuations. 
According to Ref.~\cite{Hazra2019},  the critical temperature $T_c$ is constrained by
\begin{eqnarray}
	k_B T_c \leq \frac{\pi}{2}\tilde{D},
\label{eq:contraint}
\end{eqnarray}
where  
\begin{eqnarray}
	\tilde{D} = \frac{1}{4N}\sum_{\bm{k}}\frac{\partial^2 \xi_{\bm{Q}/2+\bm{k}}}{\partial k_x^2}\langle c_{\bm{Q}/2+\bm{k}}^{\dagger}c_{\bm{Q}/2+\bm{k}}\rangle.
\end{eqnarray}
Here, $N$ denotes the number of points in the first Brillouin zone.  
We adopt the value of $\langle c_{\bm{Q}/2+\bm{k}}^{\dagger}c_{\bm{Q}/2+\bm{k}}\rangle$ for the band Hamiltonian Eq. (\ref{eq:H0}) without interaction at the same electron density. 
As an upper bound of $T_c$, $\frac{\pi}{2} \tilde{D}$ is 
particularly meaningful in the region of low values  
shown in Fig.~\ref{fig3}(e), whose shape resembles the experimentally observed high-resistivity region in Fig.~\ref{fig1}(a).
It suggests that phase fluctuations at low electron densities strongly limit $T_c$, and may even suppress superconductivity.

\begin{figure}[t]
	\begin{center}
		\includegraphics[width=8cm]{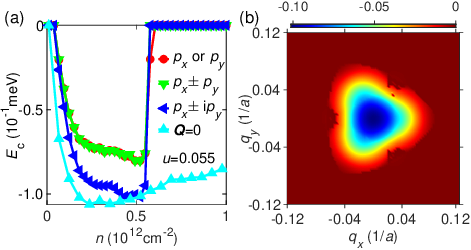}
	\end{center}
	\caption{ (a) Dependence of the condensation energy $E_c$ on electron density for total pairing momenta $\bm{Q}=2\bm{K}$ and $\bm{Q}=0$ at $u=0.055$ (eV) and $T=0.001$ (K). 
    For $\bm{Q}=2\bm{K}$, the pairing symmetries include $p_x$ or $p_y$, $p_x\pm p_y$ and $p_x\pm ip_y$. $\bm{Q}=0$ corresponds to the chiral intervalley pairing with parallel spins.
	(b)   $E_c(2\bm{K}+\bm{q})$ as a function of $\bm{q}=(q_x,q_y)$, shown on the $\bm{q}$-plane for $u=0.055$ (eV) and $n=0.5$ $(10^{12}$ cm$^{-2})$.  
			}
	\label{fig4}
\end{figure}

Below we study the stability of different pairing states by comparing their condensation energies, which can be calcuated as follows:
\begin{eqnarray}
E_c&=&\frac{1}{2}\sum_{\bm{k}}\left(|\xi_{\bm{k},\bm{Q}}|-E_{\bm{k},\bm{Q}}\right)\\
&&+\frac{\Delta_{\bm{k},\bm{Q}}\Delta_{\bm{k},\bm{Q}}^*}{8E_{\bm{k},\bm{Q}}}\left[\tanh\left(\frac{\beta E_{\bm{k},\bm{Q}}^{+}}{2}\right)+\tanh\left(\frac{\beta E_{\bm{k},\bm{Q}}^{-}}{2}\right)\right]. \nonumber
\end{eqnarray}
To mimic the transition from SC3 to SC2  as increasing $n$ in the experimental phase diagram \cite{Han2024a}, we study the evolution of the pairing states along the cut of a relatively high value of $u$, i.e., $u=0.055$ (eV).
Figure~\ref{fig4}~(a) illustrates $E_c$ as a function of $n$ for two groups of pairing states: three types of intra-valley 
pairing with $\mathbf{Q}=2\mathbf{K}$, and the inter-valley pairing with $\mathbf{Q}=0$. 
For the intra-valley pairing with $\mathbf{Q}=2\mathbf{K}$, the $p_x \pm i p_y$ state exhibits significantly lower values of $E_c$ compared to those of nematic pairing states of symmetries of $p_x$, $p_y$, and $p_x \pm p_y$, whose $E_c$'s are nearly the same. Although the inter-valley $\mathbf{Q}=0$ pairing has a much lower $E_c$ than its intra-valley counterpart, the emergence of valley polarization at large $r_s$ increases $E_c$ in the $\mathbf{Q}=0$ channel while leaving $E_c$ for $\mathbf{Q}=2\mathbf{K}$ essentially unchanged. As a result, the relative stability of different pairing states is altered. With increasing carrier density $n$, the system undergoes a transition from inter-valley $\mathbf{Q}=0$ pairing to intra-valley $\mathbf{Q}=2\mathbf{K}$ pairing. This behavior suggests that SC3 and SC2 may correspond to inter-valley and intra-valley chiral $p$-wave superconducting states, respectively.

We also check the stability of the intra-valley pairing state with $\mathbf{Q}=2\mathbf{K}$ against commensurability by calculating condensation energy $E_c$ as a function of $\bm{Q}=2\bm{K}+(q_x,q_y)$ as shown in Fig.~\ref{fig4}(b).
The results reveal that pairing with $\bm{Q}=2\bm{K}$ is energetically favored, highlighting the predominance of a commensurate pairing vector. 
Together with the calculations at two representative points corresponding to SC1 and SC2 in Fig.~\ref{fig3}(b), this emphasizes the dominance of chiral, commensurate, finite-momentum pairing in these two superconducting regions.

\begin{figure}[t]
\begin{center}
\includegraphics[width=8cm]{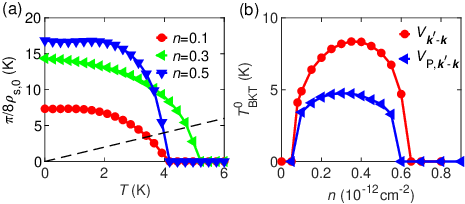}
	\end{center}
	\caption{ 
(a) Temperature dependence of the superfluid density 
$\rho_{s,0}$ at various electron densities at fixed value of $u=0.055$ (eV).
According to the Berezinskii-Kosterlitz-Thouless (BKT) theory, 
the cross points of curves with $\frac{\pi}{8}\rho_{s,0}=T_{\rm BKT}^0$ indicate 
the values of $T_{\rm BKT}^0$.
(b) Extracted BKT transition temperature for the real and complex interaction $V_{\bm{k}'-\bm{k}},~V_{P,\bm{k}'-\bm{k}}$, respectively. }
\label{fig5}
\end{figure}

In previous studies, $V_{\bm{k}^{\prime}-\bm{k}}$ in Eq. (\ref{eq:Vkk})  without the complex projection factors is used as the interaction matrix element.
We have also performed calculations by using the full expressions of $V_{P,\bm{k}'-\bm{k}}$ of Eq.  (\ref{eq:Vkk}) (see S. M. IV).
The consequential pairing symmetry remains the chiral $p$-wave. 
Nevertheless, the pairing amplitude is suppressed by the complex projection factors such that the mean-field superconducting transition temperature is further reduced by approximately one half.

Now we estimate the effect of strong phase fluctuations at low electron densities based on the Berezinskii-Kosterlitz-Thouless (BKT) theory \cite{Berezinskii1972,Kosterlitz1973,Kosterlitz1974}. 
If the superfluid density $\rho_s$ could be calculated as a function of temperature, the phase coherence temperature $T_{\rm BKT}$ can be obtained via the relation $\frac{\pi}{8}\rho_s=T_{\rm BKT}$ \cite{Nelson1977}.
However, mean-field calculations neglect fluctuation effects, leading to an overestimation of $\rho_s$ compared to the actual values. The mean-field value of $\rho_{s,0}$ \cite{Liang2017,Kitamura2022} can be calculated, as shown in Fig.~\ref{fig5}, yielding a bare BKT transition temperature $T_{BKT}^0$ that is approximately one order of magnitude larger than the experimental value. Within this framework, both the superfluid density and the phase-coherence temperature can be renormalized by a factor of roughly $1/10$--$1/5$ when the effective mass is enhanced by a factor of $5$--$10$ due to strong correlation effects. Moreover, strong interactions and phase fluctuations may further modify the form of $\rho_s$ and lead to additional suppression. A quantitative evaluation of $\rho_s$ beyond mean-field theory is therefore beyond the scope of this article.

In the experimental phase diagram, there exists another superconducting phase SC4 lying in the region of large electron density $n\approx 1.9~ (10^{12}$ cm$^{-2})$.
Our calculation also shows that in addition to the superconducting region shown in Fig. \ref{fig3}(b), superconductivity also appears at large density, for example at $n=1.9~(10^{12}$ cm$^{-2})$ and $u=0.055 $ (eV).
Different from the previous results of spin triplet pairing, self-consistent mean-field calculation shows that superconductivity only takes place for the spin-singlet inter-valley pairing.

\textit{Conclusion and Discussion--} 
In summary, we have studied the nature of the superconducting regions observed experimentally.
Based on self-consistent mean-field studies under the density-density interaction, we propose that 
SC1 and SC2 are characterized by chiral finite-momentum pairing within the same valley and spin channel, accompanied by strong phase fluctuations; SC3 corresponds to the chiral intervalley pairing. 
In contrast, SC4 features zero-momentum inter-valley spin-singlet pairing. 
Moreover, the calculated superconducting phase diagram shows qualitative agreement with experimental observations, as shown in Fig.~\ref{fig1}.

We also analyze the distribution of the dimensionless interaction parameter $r_s$ as a function of the electronic density $n$ and the displacement field $u$, where $r_s$ is defined as the ratio of the interaction energy to the kinetic energy. The cyan line in Fig.~\ref{fig1}(b) denotes the boundary of the $r_s > 40$ region. In this regime, strong-correlation effects may become significant and induce Wigner crystal or competing ordered instabilities \cite{Kim2022a,Attaccalite2002,Drummond2009}, which could separate the SC1 and SC2 phases. In most of the superconducting region, $r_s$ lies between 10 and 30, which may render vertex corrections important and substantially reduce $T_c$ \cite{Huang2006,Tazai2016}.

Experimentally \cite{Han2024a}, the observation of hysteretic loops in magnetic-field-dependent resistivity measurements, along with well-defined quantum oscillations, suggests the emergence of a spin-valley-polarized metallic state. Accompanying anomalous Hall signals further confirm the breaking of time-reversal symmetry. This spin–valley–polarized metallic state at low electron density is further corroborated by theoretical analyses of spin–valley instabilities within the RPA framework \cite{Parra-Martinez2025}. In addition, the chiral finite-momentum pairing between electrons in the same valley and with the same spin in the SC1 and SC2 phases may give rise to a phase-fluctuation-induced higher-order time-reversal-symmetry-breaking normal state. The interplay between such unconventional pairing and phase fluctuations has also been proposed to give rise to a variety of exotic phases and novel effects.
\cite{Wu2005a,Berg2009,Wu2010,Zeng2021a,Liu2023,Maccari2022, Qin2023PRB,Qin2024,Jian2021,Fernandes2021,Ge2024,
Grinenko2021b,Pan2024,Orazbay2024}.

\textit{ Note added}--During the preparation of the manuscript, 
we became aware of a related work \cite{Yang2024c}.

This work was supported by National Natural Science Foundation of China (Grants No. 12447125, No. 12234016, No. 12174317) and the New Cornerstone Science Foundation.

\onecolumngrid

\newpage
\onecolumngrid
\begin{appendix}
	\newpage
	\begin{center} {\large \textbf{Chiral finite-momentum superconductivity in the tetralayer graphene\\
			\vspace{0.2cm}
			- Supplemental Material -}}
	\end{center}
	\renewcommand{\thefigure}{S\arabic{figure}}
	\renewcommand{\theequation}{S\arabic{equation}}
	\setcounter{equation}{0}
	\setcounter{figure}{0}
	
	\setcounter{table}{0}
	
\section{I. Energy Dispersion}
The non-interacting part of the Hamiltonian is given by
\begin{eqnarray}
	H_0=\sum_{\bm{k}s\tau}\Psi_{\bm{k}\tau s}^{\dagger}h_{\bm{k}\tau}\Psi_{\bm{k}\tau s},
\end{eqnarray}
where $s$ denotes the spin index and $\tau$ represents the valley index. The field operator $\Psi_{\bm{k}\tau s}^{\dagger}=\left(\psi_{\bm{k}\tau s,1A}^{\dagger},~\cdots, ~\psi_{\bm{k}\tau s, 4B}^{\dagger} \right)$ consists of creation operators $\psi_{\bm{k}\tau s,l\sigma}^{\dagger}$ at the wavevector $\bm{k}$, where $l=1,2,3,4$ denotes the layer index and $\sigma=A,B$ represents the sublattice index. Following previous works \cite{Zhou2021c, Chou2021a, Ghazaryan2021b, Qin2023c, Han2024a}, we adopt the following expression for the matrix $h_{\bm{k}\tau}$:
\begin{eqnarray}
	h_{\bm{k}\tau}=\left(\begin{array}{cccccccc}
		u&v_0\pi^{\dagger}&v_4\pi^{\dagger}&0&0&\gamma/2&0&0\\
		v_0\pi&u&\gamma_1&v_4\pi^{\dagger}&0&0&0&0\\
		v_4\pi&\gamma_1&u/3&v_0\pi^{\dagger}&v_4\pi^{\dagger}&v_3\pi&0&\gamma_2/2\\
		v_3\pi^{\dagger}&v_4\pi&v_0\pi&u/3&\gamma_1&v_4\pi^{\dagger}&0&0\\
		0&0&v_4\pi&\gamma_1&-u/3&v_0\pi^{\dagger}&v_4\pi^{\dagger}&v_3\pi\\
		\gamma_2/2&0&v_3\pi^{\dagger}&v_4\pi&v_0\pi&-u/3&\gamma_1&v_4\pi^{\dagger}\\
		0&0&0&0&v_4\pi&\gamma_1&-u&v_0\pi^{\dagger}\\
		0&0&\gamma_2/2&0&v_3\pi^{\dagger}&v_4\pi&v_0\pi&-u
	\end{array}\right).
\end{eqnarray}
In this model, $v_i=\frac{\sqrt{3}}{2}\gamma_i$ and $\pi=\tau k_x+ik_y$, where the valley index $\tau=\pm1$. The parameters $\gamma_0=3.1$ eV, $\gamma_1=0.38$ eV, $\gamma_2=-0.015$ eV, $\gamma_3=-0.29$ eV, and $\gamma_4=-0.141$ eV are used in the Hamiltonian. Additionally, $u$ is proportional to the displacement field. For the calculation of the electron density, the lattice constant $a=0.246$ nm was adopted. 

\begin{figure}[htbp]
	\begin{center}
		\includegraphics[width=16cm]{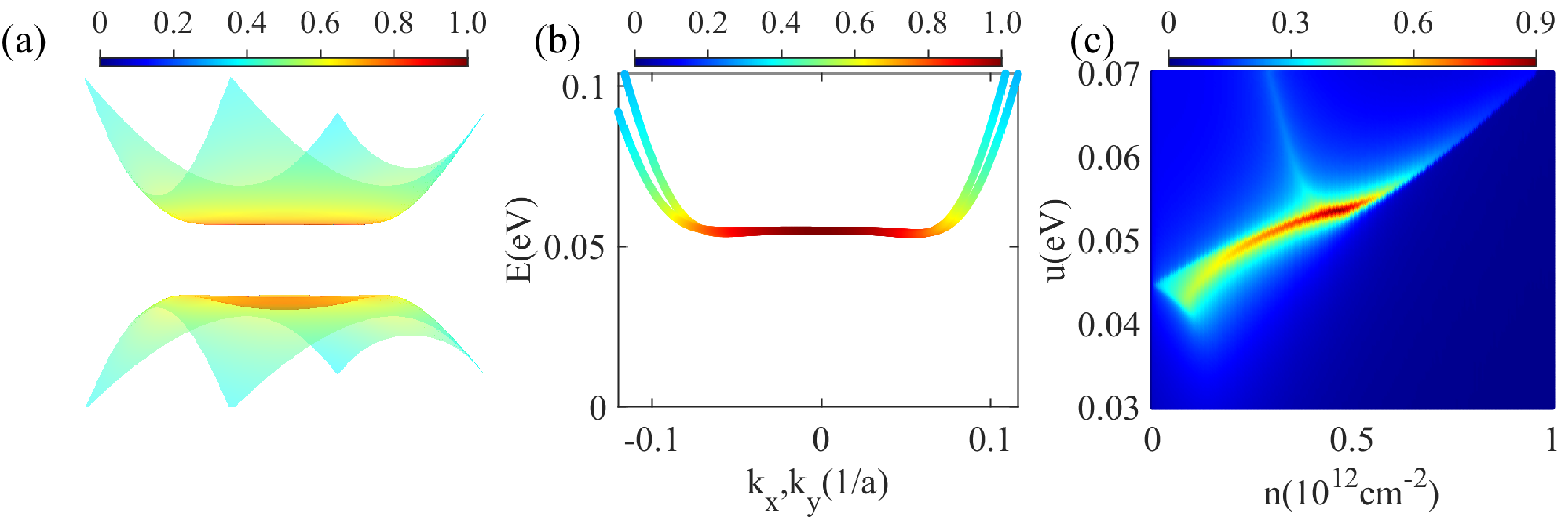}
	\end{center}
	\caption{ (a) The energy dispersion at the displacement field $u=0.055$ eV, where the color represents the orbital weight of 1A and 4B. (b) The energy dispersion along the $k_x$ and $k_y$ direction at $u=0.055$ eV, where color indicates the orbital weight of 1A. (c) The distribution of density of state $N(E_F)$(meV$^{-1}$) across different dispalcement field $u$ and  electron density $n$.}
	\label{figS1}
\end{figure}

In our tight-binding model calculations, we observed that the orbital weight of the electrons at sites $1A$ and $4B$ predominantly contributes to the first conduction and valence bands, as shown in Fig. \ref{figS1}(a). For the conduction band, the electron density at the $1A$ site is dominant, as illustrated in Fig. \ref{figS1}(b). The density of states as a function of the displacement field $u$ and electron density $n$ is presented in Fig. \ref{figS1}(c).

\section{II. Interaction}
The density-density interaction can be expressed as:
\begin{eqnarray}
	H_I=\frac{1}{2A}\sum_{\bm{q}}V_{0,\bm{q}}\rho_{\bm{q}}\rho_{-\bm{q}},
\end{eqnarray}
where the density operator is defined as $\rho_{\bm{q}}=\sum_{\bm{k}\tau s}\Psi_{\bm{k}\tau s}^{\dagger}\Psi_{\bm{k}+\bm{q}\tau s}$, $A$ is the 2D area, and the Coulomb interaction is given by $V_{0,\bm{q}}=\frac{e^2}{2\epsilon q}\tanh(qd)$ \cite{Ghazaryan2021b,Guerci2024}.

By diagonalizing the non-interacting part for each spin, valley, and momentum, we obtain $\Psi^{\dagger}_{\bm{k}\tau s}h_{\bm{k}\tau}\Psi_{\bm{k}\tau s}=\varphi^{\dagger}_{\bm{k}\tau s}\Lambda_{\bm{k}\tau}\varphi_{\bm{k}\tau s}$. This implies that $\Psi_{\bm{k}\tau s}=U_{\bm{k}\tau s}\varphi_{\bm{k}\tau s}$, or equivalently, $\psi_{\bm{k}\tau s,i}=\sum_{\mu}u_{\tau,i\mu}(\bm{k})c_{\bm{k}\tau s,\mu}$. With this transformation, the density operator becomes $\rho_{\bm{q}}=\sum_{\bm{k}\tau si}\sum_{\mu\mu'}\left(u_{\tau,i\mu}^*(\bm{k})u_{\tau,i\mu'}(\bm{k}+\bm{q})\right)c_{\bm{k}\tau s,\mu}^{\dagger}c_{\bm{k}+\bm{q}\tau s,\mu'}$.
This leads to a density operator projected onto the first conduction band $\mu_0$, given by
\begin{eqnarray}
	\tilde{\rho}_{\bm{q}}=\sum_{\bm{k}\tau s}\sum_i\left(u_{\tau,i\mu_0}^*(\bm{k})u_{\tau,i\mu_0}(\bm{k}+\bm{q})\right)c_{\bm{k}\tau s,\mu_0}^{\dagger}c_{\bm{k}+\bm{q}\tau s,\mu_0}.
\end{eqnarray}

Thus, the interaction takes the following form:
\begin{eqnarray}
	H_I&=&\frac{1}{2A}\sum_{\bm{q}}V_{0,\bm{q}}\tilde{\rho}_{\bm{q}}\tilde{\rho}_{-\bm{q}}\\
	&=&\frac{1}{2A}\sum_{\bm{q}}V_{0,\bm{q}}\left[\sum_{\bm{k}\tau s}\sum_iu_{\tau,i\mu_0}^*(\bm{k})u_{\tau,i\mu_0}(\bm{k}+\bm{q})c_{\bm{k}\tau s,\mu_0}^{\dagger}c_{\bm{k}+\bm{q}\tau s,\mu_0}
	\right]\nonumber\\
	&\times&
	\left[\sum_{\bm{k}'\tau' s'}\sum_ju_{\tau',j\mu_0}^*(\bm{k}')u_{\tau',j\mu_0}(\bm{k}'-\bm{q})c_{\bm{k}'\tau' s',\mu_0}^{\dagger}c_{\bm{k}'-\bm{q}\tau' s',\mu_0}
	\right].
\end{eqnarray}

In the random phase approximation (RPA), the interaction is modified as $V_{\bm{q}}=\frac{V_{0,\bm{q}}}{1+\chi_{0,\bm{q}}V_{0,\bm{q}}}$,
where we have neglected the frequency dependence of the  density-density susceptibility $\chi_{0,\bm{q}}$. It can then be written as:
\begin{eqnarray}
	\chi_0(\bm{q},\tau)&=&\langle T_{\tau}\left(\tilde{\rho}_{\bm{q}}(\tau)\tilde{\rho}_{\bm{q}}(0)\right)\rangle\nonumber\\
	&=&-\sum_{\bm{k}\tau s}\sum_{ij}\left(u_{\tau,i\mu_0}^*(\bm{k})u_{\tau,i\mu_0}(\bm{k}+\bm{q})\right)\left(u_{\tau,j\mu_0}^*(\bm{k}+\bm{q})u_{\tau,j\mu_0}(\bm{k})\right)G_{0}^{\tau s,\mu_0}(\bm{k}+\bm{q},\tau)G_{0}^{\tau s,\mu_0}(\bm{k},-\tau).
\end{eqnarray}
By defining 
\begin{eqnarray}
	g(\bm{k},\bm{q},\tau)=-G_{0}^{\tau s,\mu_0}(\bm{k}+\bm{q},\tau)G_{0}^{\tau s,\mu_0}(\bm{k},-\tau),
\end{eqnarray}
we obtian
\begin{eqnarray}
	g(\bm{k},\bm{q},i\Omega_n)&=&\frac{-1}{\beta}\sum_{\omega_n}G_{0}^{\tau s,\mu_0}(\bm{k}+\bm{q},i\Omega_n+i\omega_n)G_{0}^{\tau s,\mu_0}(\bm{k},i\omega_n)=\frac{f(\xi_{\bm{k}+\bm{q}}^{\tau s,\mu_0})-f(\xi_{\bm{k}}^{\tau s,\mu_0})}{i\Omega_n+\xi_{\bm{k}}^{\tau s,\mu_0}-\xi_{\bm{k}+\bm{q}}^{\tau s,\mu_0}},
\end{eqnarray}
where $f(\xi_{\bm{k}}^{\tau s,\mu_0})$ denotes the Fermi-Dirac distribution function. This leads to the following expression:
\begin{eqnarray}
	\chi_0(\bm{q},i\Omega_n)&=&\sum_{\bm{k}\tau s}\left|\left[\sum_{i}u_{\tau,i\mu_0}^*(\bm{k})u_{\tau,i\mu_0}(\bm{k}+\bm{q})\right]\right|^2\frac{f(\xi_{\bm{k}+\bm{q}}^{\tau s,\mu_0})-f(\xi_{\bm{k}}^{\tau s,\mu_0})}{i\Omega_n+\xi_{\bm{k}}^{\tau s,\mu_0}-\xi_{\bm{k}+\bm{q}}^{\tau s,\mu_0}}.
\end{eqnarray}

The interaction term takes the following form:
\begin{eqnarray}
	H_I&=&\frac{1}{2A}\sum_{\bm{q}}V_{\bm{q}}\left[\sum_{\bm{k}}\sum_{i\tau s}\left(u_{\tau,i\mu_0}^*(\bm{k})u_{\tau,i\mu_0}(\bm{k}+\bm{q})\right)c_{\bm{k}\tau s,\mu_0}^{\dagger}c_{\bm{k}+\bm{q}\tau s,\mu_0}
	\right]
	\nonumber\\&\times&\left[\sum_{\bm{k}'}\sum_{j\tau's'}\left(u_{\tau',j\mu_0}^*(\bm{k}')u_{\tau',j\mu_0}(\bm{k}'-\bm{q})\right)c_{\bm{k}'\tau s,\mu_0}^{\dagger}c_{\bm{k}'-\bm{q}\tau s,\mu_0}
	\right]\nonumber\\
	&=&\frac{1}{2A}\sum_{\bm{q}\bm{k}\bm{k}'}\sum_{ss'}\sum_{\tau\tau'}V_{P,\bm{q}}^{\tau\tau'}c_{\bm{k}\tau s,\mu_0}^{\dagger}c_{\bm{k}+\bm{q}\tau s,\mu_0}c_{\bm{k}'\tau s,\mu_0}^{\dagger}c_{\bm{k}'-\bm{q}\tau s,\mu_0},
\end{eqnarray}
where the interaction strength $V_{P,\bm{q}}^{\tau\tau'}$ can be written as:
\begin{eqnarray}
	V_{P,\bm{q}}^{\tau\tau'}=V_{\bm{q}}\left[\sum_{i}u_{\tau,i\mu_0}^*(\bm{k})u_{\tau,i\mu_0}(\bm{k}+\bm{q})\right]\left[\sum_{j}\left(u_{\tau',j\mu_0}^*(\bm{k}')u_{\tau',j\mu_0}(\bm{k}'-\bm{q})\right)\right].
\end{eqnarray}

When considering the pairing interaction only between states with the same spin $s=s_0$ and the same valley $\tau=\tau_0$, the interaction term is given by
\begin{eqnarray}
	H_I&=&\frac{1}{2A}\sum_{\bm{q}\bm{k}\bm{k}'}V_{P,\bm{q}}c_{\bm{k}}^{\dagger}c_{\bm{k}+\bm{q}}c_{\bm{k}'}^{\dagger}c_{\bm{k}'-\bm{q}}=\frac{1}{2A}\sum_{\bm{k}'\bm{k}\bm{Q}}V_{P,\bm{k}'-\bm{k}}c_{\bm{Q}/2+\bm{k}}^{\dagger}c_{\bm{Q}/2-\bm{k}}^{\dagger}c_{\bm{Q}/2-\bm{k}'}c_{\bm{Q}/2+\bm{k}'}.
\end{eqnarray}
Here, we omit the indices $s_0, \tau_0,\mu_0$, and the interaction potential $V_{P,\bm{k}'-\bm{k}}$ is expressed as:
\begin{eqnarray}
	V_{P,\bm{k}'-\bm{k}}
	&=&V_{\bm{k}'-\bm{k}}\left[\sum_{i}u_{i\mu_0}^*(\bm{Q}/2+\bm{k})u_{i\mu_0}(\bm{Q}/2+\bm{k}')\right]\left[\sum_ju_{j\mu_0}^*(\bm{Q}/2-\bm{k})u_{j\mu_0}(\bm{Q}/2-\bm{k}')\right].
\end{eqnarray}

We also note that the interaction in real space exhibits an oscillatory form. The position of the locally maximal attractive interaction suggests that pairing occurs at distances about 10 nm, which is comparable to the electrons separation.

\section{III. Mean Field Analysis}
The Hamiltonian for finite momentum pairing with the same spin and valley can be written as:
\begin{equation}
	H=\sum\limits_{\bm{k}} \xi_{\bm{k}} c^{\dagger}_{\bm{k}} c_{\bm{k}}+\sum\limits_{\bm{k}, \bm{k}'} V_{\bm{k},\bm{k}'}c^{\dagger}_{\bm{Q}/2+\bm{k}}
	c^{\dagger}_{\bm{Q}/2-\bm{k}}c_{\bm{Q}/2-\bm{k}'}c_{\bm{Q}/2+\bm{k}'}.
\end{equation}
In the mean field approximation, the interaction term becomes
\begin{eqnarray}
	\sum\limits_{\bm{k}, \bm{k}'} V_{\bm{k},\bm{k}'}c^{\dagger}_{\bm{Q}/2+\bm{k}}
	c^{\dagger}_{\bm{Q}/2-\bm{k}}c_{\bm{Q}/2-\bm{k}'}c_{\bm{Q}/2+\bm{k}'}\rightarrow -\frac{1}{2}\sum_{\bm{k}}\Delta^*_{\bm{k},\bm{Q}}c_{\bm{Q}/2-\bm{k}}c_{\bm{Q}/2+\bm{k}}
	-\frac{1}{2}\sum_{\bm{k}}\Delta_{\bm{k},\bm{Q}}c^{\dagger}_{\bm{Q}/2+\bm{k}}c^{\dagger}_{\bm{Q}/2-\bm{k}}+{\rm const},
\end{eqnarray}
where we define $\Delta_{\bm{k},\bm{Q}}=-2\sum_{\bm{k}'}V_{\bm{k},\bm{k}'}\langle c_{\bm{Q}/2-\bm{k}'}c_{\bm{Q}/2+\bm{k}'}\rangle$. When using Nambu spinor $\psi_{\bm{k},\bm{Q}}^{\dagger}=\left(
c^{\dagger}_{\bm{Q}/2+\bm{k}}~c_{\bm{Q}/2-\bm{k}} \\
\right)$, the mean field hamitonain becomes
\begin{eqnarray}
	H=\sum_{\bm{k}}^{\prime}\psi_{\bm{k},\bm{Q}}^{\dagger}\hat{h}_{\bm{k},\bm{Q}}\psi_{\bm{k},\bm{Q}}+{\rm const},
\end{eqnarray}
where $\sum^{\prime}$ represents the summation over half of the $\bm{k}$ grid and
\begin{eqnarray}
	\hat{h}_{\bm{k},\bm{Q}}=\left(\begin{matrix}
		\xi_{\bm{Q}/2+\bm{k}}&-\Delta_{\bm{k},\bm{Q}}\\
		-\Delta_{\bm{k},\bm{Q}}^*&-\xi_{\bm{Q}/2-\bm{k}}
	\end{matrix}\right).
\end{eqnarray}
By diagonalizing process, we know $\psi_{\bm{k},\bm{Q}}^{\dagger}=\Gamma_{\bm{k},\bm{Q}}^{\dagger}U_{\bm{k},\bm{Q}}^{\dagger}$, where $\Gamma_{\bm{k},\bm{Q}}^{\dagger}=\left(
\gamma_{\bm{Q}/2+\bm{k}}^{\dagger}~\gamma_{\bm{Q}/2-\bm{k}}
\right)$ and unitary matrix $U_{\bm{k},\bm{Q}}=\left(
\begin{matrix}
	u_{\bm{k},\bm{Q}} & v_{\bm{k},\bm{Q}}\\
	-v^*_{\bm{k},\bm{Q}} & u^*_{\bm{k},\bm{Q}}\\
\end{matrix}
\right)$. Here $|u_{\bm{k},\bm{Q}}|^2=\frac{1}{2}(1+\frac{\xi_{\bm{k},\bm{Q}}}{E_{\bm{k},\bm{Q}}})$,  $|v_{\bm{k},\bm{Q}}|^2=\frac{1}{2}(1-\frac{\xi_{\bm{k},\bm{Q}}}{E_{\bm{k},\bm{Q}}})$, and $	u_{\bm{k},\bm{Q}}v_{\bm{k},\bm{Q}}=\frac{\Delta_{\bm{k},\bm{Q}}}{2E_{\bm{k,\bm{Q}}}}$, where $\xi_{\bm{k},\bm{Q}}=\frac{1}{2}(\xi_{\bm{Q}/2+\bm{k}}+\xi_{\bm{Q}/2-\bm{k}})$ and $E_{\bm{k},\bm{Q}}=\sqrt{\xi_{\bm{k},\bm{Q}}^2+|\Delta_{\bm{k},\bm{Q}}|^2}$.
These calculations give the following gap equation as: 
\begin{eqnarray}
	\Delta_{\bm{k},\bm{Q}}=-\sum_{\bm{k}'}V_{\bm{k},\bm{k}'}\left(\frac{\Delta_{\bm{k}',\bm{Q}}}{2E_{\bm{k}',\bm{Q}}}\right)\left[\tanh\left(\frac{\beta E_{\bm{k},\bm{Q}}^{+}}{2}\right)+\tanh\left(\frac{\beta E_{\bm{k},\bm{Q}}^{-}}{2}\right)\right],
\end{eqnarray}
where $E_{\bm{k},\bm{Q}}^{\pm}=\pm \frac{1}{2}\left(\xi_{\bm{Q}/2+\bm{k}}-\xi_{\bm{Q}/2-\bm{k}}\right)+E_{\bm{k},\bm{Q}}$.
Similarly, the condensation energy can be written as:
\begin{eqnarray}
	E_c=\frac{1}{2}\sum_{\bm{k}}\left(|\xi_{\bm{k},\bm{Q}}|-E_{\bm{k},\bm{Q}}\right)+\frac{\Delta_{\bm{k},\bm{Q}}\Delta_{\bm{k},\bm{Q}}^*}{8E_{\bm{k},\bm{Q}}}\left[\tanh\left(\frac{\beta E_{\bm{k},\bm{Q}}^{+}}{2}\right)+\tanh\left(\frac{\beta E_{\bm{k},\bm{Q}}^{-}}{2}\right)\right].
\end{eqnarray}

For the pairing interaction of different valley, the gap equation takes the following form:
\begin{eqnarray}
	\Delta_{\bm{k}}=-\sum_{\bm{k}'}V_{\bm{k},\bm{k}'}\left(\frac{\Delta_{\bm{k}'}}{2E_{\bm{k}}}\right)\tanh\left(\frac{\beta E_{\bm{k}}}{2}\right),
\end{eqnarray}
where $E_{\bm{k}}=\sqrt{\xi_{\bm{k}}^2+|\Delta_{\bm{k}}|^2}$. The condensation energy has the following form:
\begin{eqnarray}
	E_c=\sum_{\bm{k}}\left[|\xi_{\bm{k}}|-E_{\bm{k}}+\frac{\Delta_{\bm{k}}\Delta_{\bm{k}}^*}{2E_{\bm{k}}}\tanh\left(\frac{\beta E_{\bm{k}}}{2}\right)\right].
\end{eqnarray}

The superfluid density, denoted as $\rho_s$, can also be determined and is formally defined as:
\begin{eqnarray}
	\rho_s=\langle -K_x\rangle-\Pi_x(q_x=0,q_y\rightarrow0,i\Omega_n=0),
\end{eqnarray}
where  
\begin{eqnarray}
	\langle -K_x\rangle&=&
	\sum_{\bm{k}}\left[\langle c_{\bm{Q}/2+\bm{k}}^{\dagger}c_{\bm{Q}/2+\bm{k}}\rangle\left(\frac{\partial^2 \xi_{\bm{Q}/2+\bm{k}}}{\partial k_x^2}\right)\right]\nonumber\\
	&=&\sum_{\bm{k}}\left[\frac{1}{2}(1+\frac{\xi_{\bm{k},\bm{Q}}}{E_{\bm{k},\bm{Q}}})f(E_{\bm{k},\bm{Q}}^+)+\frac{1}{2}(1-\frac{\xi_{\bm{k},\bm{Q}}}{E_{\bm{k},\bm{Q}}})(1-f(E_{\bm{k},\bm{Q}}^-))\right]\left(\frac{\partial^2 \xi_{\bm{Q}/2+\bm{k}}}{\partial k_x^2}\right).
\end{eqnarray}
And $\Pi_x(\bm{q},\tau)$ represents the current-current correlation function, defined as: 
\begin{eqnarray}
	\Pi_{x}(\bm{q},\tau)&=&\left\langle T_{\tau}\left[j_x(\bm{q},\tau)j_x(-\bm{q},0)\right]\right\rangle,
\end{eqnarray}
where 
\begin{eqnarray}
	j_x(\bm{q},0)=e^{iq_x/2}\sum_{\bm{k}\sigma}v_{\bm{Q}/2+\bm{k}+\bm{q}/2}^xc_{\bm{Q}/2+\bm{k}}^{\dagger}c_{\bm{Q}/2+\bm{k}+\bm{q}},~~~v_{\bm{Q}/2+\bm{k}}^x=\frac{\partial \xi_{\bm{Q}/2+\bm{k}}}{\partial k_x}.
\end{eqnarray}
By applying the relation $	\Pi_{x}(\bm{q},i\Omega_n)=\int_{0}^{\beta}d\tau\Pi_{x}(\bm{q},\tau)e^{i\Omega_n\tau}$, the current-current correlation function can be expressed as:
\begin{eqnarray}
	\Pi_{x}(\bm{q},i\Omega_n)&=&\frac{1}{\beta}\sum_{i\omega_n}\sum_{\bm{k}}\left[-(v_{\bm{Q}/2+\bm{k}+\bm{q}/2}^x)^2G(\bm{k}+\bm{q},i\omega_n+i\Omega_n)G(\bm{k},i\omega_n)\right.\nonumber\\
	&&\left.+v_{\bm{Q}/2+\bm{k}+\bm{q}/2}v_{\bm{Q}/2-\bm{k}-\bm{q}/2}F(\bm{k}+\bm{q},i\omega_n+i\Omega_n)\bar{F}(\bm{k},i\omega_n)\right],
\end{eqnarray}
where $G(\bm{k},i\omega_n)$ and $F(\bm{k},i\omega_n)$
denote the normal and anomalous Green's functions, respectively. They have the following form:
\begin{eqnarray}
	G(\bm{k},i\omega_n)&=&\frac{i\omega_n+\xi_{\bm{Q}/2-\bm{k}}}{(i\omega_n-E_{\bm{k},\bm{Q}}^+)(i\omega_n+E_{\bm{k},\bm{Q}}^-)},\\
	F(\bm{k},i\omega_n)&=&\frac{-\Delta_{\bm{k},\bm{Q}}}{(i\omega_n-E_{\bm{k},\bm{Q}}^+)(i\omega_n+E_{\bm{k},\bm{Q}}^-)},~\bar{F}(\bm{k},i\omega_n)=\frac{-\Delta_{\bm{k},\bm{Q}}^*}{(i\omega_n-E_{\bm{k},\bm{Q}}^+)(i\omega_n+E_{\bm{k},\bm{Q}}^-)}.
\end{eqnarray} 

Thus 
\begin{eqnarray}
	\Pi_x(q_x=0,q_y\rightarrow0,0)&=&\frac{1}{4}\sum_{\bm{k}}\left[-(v_{\bm{Q}/2+\bm{k}}^x)^2(1+\frac{\xi_{\bm{k},\bm{Q}}^2}{E_{\bm{k},\bm{Q}}^2})+v_{\bm{Q}/2+\bm{k}}^xv_{\bm{Q}/2-\bm{k}}^x\frac{|\Delta_{\bm{k},\bm{Q}}|^2}{E_{\bm{k},\bm{Q}}^2}\right]\left(\frac{\partial f}{\partial E_{\bm{k},\bm{Q}}^+}+\frac{\partial f}{\partial E_{\bm{k},\bm{Q}}^-}\right)\nonumber\\
	&+&\frac{1}{2}\sum_{\bm{k}}\left[(v_{\bm{Q}/2+\bm{k}}^x)^2(1-\frac{\xi_{\bm{k},\bm{Q}}^2}{E_{\bm{k},\bm{Q}}^2})+v_{\bm{Q}/2+\bm{k}}^xv_{\bm{Q}/2-\bm{k}}^x\frac{|\Delta_{\bm{k},\bm{Q}}|^2}{E_{\bm{k},\bm{Q}}^2}\right]\frac{f(E_{\bm{k},\bm{Q}}^+)+f(E_{\bm{k},\bm{Q}}^-)-1}{E_{\bm{k},\bm{Q}}^++E_{\bm{k},\bm{Q}}^-}\nonumber\\
	&-&\frac{1}{2}\sum_{\bm{k}}\frac{\xi_{\bm{k},\bm{Q}}}{E_{\bm{k},\bm{Q}}}(v_{\bm{Q}/2+\bm{k}}^x)^2\left(\frac{\partial f}{\partial E_{\bm{k},\bm{Q}}^+}-\frac{\partial f}{\partial E_{\bm{k},\bm{Q}}^-}\right).
\end{eqnarray}

\begin{figure}[htbp]
	\begin{center}
		\includegraphics[width=16cm]{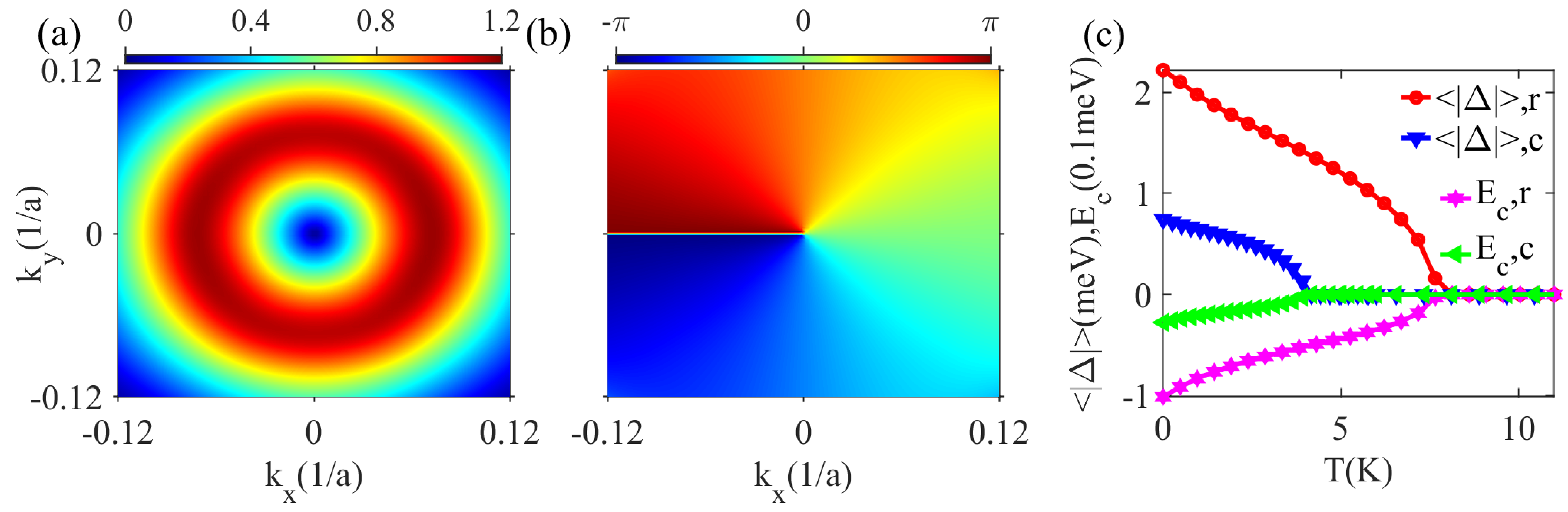}
	\end{center}
	\caption{ 
		(a) Momentum dependence of the amplitude and (b) phase of $\Delta_{\bm{k}\bm{Q}}$ at $u=0.055$ (eV) and $n=0.5$ $(10^{12}$ cm$^{-2})$ for the complex interaction.  
		(c) Temperature dependence of the average absolute  value of the pairing gap, $\langle|\Delta|\rangle$, and the condensation energy, $E_c$, for real (r) and complex (c) interactions.
	}
	\label{figS2}
\end{figure}

\section{IV. Superconducting gap for complex interaction}

In addition to the calculations about the real interaction $V_{\bm{k}'-\bm{k}}$, we also solve the gap equation self-consistently for the complex interaction $V_{P,\bm{k}'-\bm{k}}$. Figures \ref{figS2}(a) and \ref{figS2}(b) illustrate the momentum dependence of the pairing gap at $u=0.055$ (eV) and $n=0.5$ $(10^{12}$ cm$^{-2})$, revealing a chiral $p$-wave pairing symmetry. However, when the gauge choice is altered, the gap symmetry also changes, with the winding number of the dominant phase factor playing a contributing role.
The temperature dependence of the average absolute value of the pairing gap, $\langle|\Delta|\rangle$, and the condensation energy $E_c$ in Fig.~\ref{figS2}(c) demonstrates that the mean-field superconducting transition temperature is further reduced by about half under the complex interaction compared to the real interaction.

\end{appendix}

\end{document}